\documentclass[12pt,thmsa]{article}%
\usepackage{amssymb}
\usepackage{sw20lart}
\usepackage{amsmath}
\usepackage{amsfonts}
\usepackage{graphicx}%
\begin{document}

\title{Physical Consonance Law of Sound Waves }
\author{Mario Goto\\(mgoto@uel.br)\\Departamento de F\'{\i}sica\\Centro de Ci\^{e}ncias Exatas \\Universidade Estadual de Londrina}
\date{December 18, 2004 Revised: June 14, 2005}
\maketitle

\begin{abstract}
Sound consonance is the reason why it is possible to exist music in our life.
However, rules of consonance between sounds had been found quite subjectively,
just by hearing. To care for, the proposal is to establish a sound consonance
law on the basis of mathematical and physical foundations. Nevertheless, the
sensibility of the human auditory system to the audible range of frequencies
is individual and depends on a several factors such as the age or the health
in a such way that the human perception of the consonance as the pleasant
sensation it produces, while reinforced by an exact physical relation, may
involves as well the individual subjective feeling

\end{abstract}

\section{\smallskip Introduction}

Sound consonance is one of the main reason why it is possible to exist music
in our life. However, rules of sound consonance had been found quite
subjectively, just by hearing \cite{Thomas}. It sounds good, after all music
is art! But physics challenge is to discover laws wherever they are
\cite{Feynman}. To care for, it is proposed, here, a sound consonance law on
the basis of mathematical and physical foundations.

As we know, Occidental musics are based in the so called Just Intonation
Scale, built with a set of musical notes which frequencies are related between
them in the interval of frequencies from some $f_{0}$ to $f_{1}=2f_{0}$ that
defines the octave. Human audible sound frequencies comprehend from about
$20Hz$ to $20,000Hz$, and a typical piano keyboard covers 7 octaves from notes
$A_{0}$ to $C_{8}$, with frequencies $27.5Hz$ and $4,224Hz$, respectively,
using the standard tuning up frequency attributed to the note $A_{4}$
($440Hz$) \cite{Harry}, \cite{Lapp}$.$

The origin of actual musical scale remits us to the Greek mathematician
Pythagoras. Using a monochord, a vibrating string with a movable bridge that
transforms into a two vibrating strings with different but related lengths and
frequencies, he found that combinations of two sounds with frequency relations
$2:1$, $3:2$ and $4:3$ are particularly pleasant, while many other arbitrary
combinations are unpleasant. Two sounds getting a pleasant combined sound are
called consonants, otherwise they are dissonant. This set of relations
encompassed into the frequencies interval of one octave defines the
Pythagorean Scale, which is shown in Table 1, with notes and frequency
relations of each note compared to the first one, $C_{i}$. Table 2 shows the
frequency relations of the two adjacent notes, with a tone given by 9/8 and a
semitone given by 256/243.

\smallskip

\begin{center}%
\begin{tabular}
[c]{|c|c|c|c|c|c|c|c|}\hline
$C_{i}$ & $D$ & $E$ & $F$ & $G$ & $A$ & $B$ & $C_{f}$\\\hline
$1$ & $9/8$ & $81/64$ & $4/3$ & $3/2$ & $27/16$ & $243/128$ & $2$\\\hline
\end{tabular}

Table 1: notes and frequency relations compared with the first note $C_{i}$,
in the Pythagorean scale.

\bigskip%

\begin{tabular}
[c]{|c|c|c|c|c|c|c|}\hline
$D/C_{i}$ & $E/D$ & $F/E$ & $G/F$ & $A/G$ & $B/A$ & $C_{f}/B$\\\hline
9/8 & 9/8 & 256/243 & 9/8 & 9/8 & 9/8 & 256/243\\\hline
\end{tabular}

Table 2: notes and frequency relations of the two consecutive notes, in the
Pythagorean scale.
\end{center}

\bigskip

Consonance and dissonance are not absolute concepts, and the Greek astronomer
Ptolomy added to the Pythagorean consonant relations $3:2:1$ another set of
relations who considered as well as consonant, $4:5:6$. This enlarged set is
the base of the so called Just Intonation Scale, which is shown in table 3,
with notes and frequency relations of each note compared to the first one,
$C_{i}$. Table 4 shows the frequency relations of the two adjacent notes, with
a major tone given by 9/8, the minor tone by 10/9 and a semitone by 16/15.

\begin{center}
\smallskip%

\begin{tabular}
[c]{|c|c|c|c|c|c|c|c|}\hline
$C_{i}$ & $D$ & $E$ & $F$ & $G$ & $A$ & $B$ & $C_{f}$\\\hline
$1$ & $9/8$ & $5/4$ & $4/3$ & $3/2$ & $5/3$ & $15/8$ & $2$\\\hline
\end{tabular}

Table 3: notes and frequency relations compared with the first note $C_{i}$,
in the Just Intonation Scale.

\bigskip%

\begin{tabular}
[c]{|c|c|c|c|c|c|c|}\hline
$D/C_{i}$ & $E/D$ & $F/E$ & $G/F$ & $A/G$ & $B/A$ & $C_{f}/B$\\\hline
$9/8$ & $10/9$ & $16/15$ & $9/8$ & $10/9$ & $9/8$ & $16/15$\\\hline
\end{tabular}

Table 4: notes and frequency relations of the two consecutive notes in the
Just Intonation Scale.
\end{center}

\bigskip

Actually, in terms of the consonant frequency relations, the Just Intonation
Scale is richer than the Pythagorean Scale. It can be understood why it is
just so attempting to the physical condition to the sound waves consonance we
are going to establish in next section.

Notice that all natural sound can be treated as composed by a linear
combination of harmonic waves with frequencies related to the fundamental one,
$f_{0}$, as $f_{n}=nf_{0}$ for integer $n$ $(=1,2,3,...)$. All these harmonic
components are considered as consonant with the fundamental and the relative
weight of these harmonics is a characteristic of the sound source, which
frequencies spectrum defines one of the most important sound quality, the timbre.

Natural musical scale as the Just Intonation Scale implies some practical
problems due to the unequal frequency relations that define one tone or half
tone related notes. To avoid such trouble, it was created the scale of equal
temperament (chromatic scale), composed with 12 notes with equal frequency
relation $r$ between adjacent notes from a fundamental $f_{0}$ to the octave
above $f_{12}=r^{12}f_{0}=2f_{0}$ frequencies,
\begin{equation}
f_{0},\ f_{1}=rf_{0,}\ f_{2}=rf_{1}=r^{2}f_{0},\ f_{3}=r^{3}f_{0}%
,\ \cdots,\ f_{12}=r^{12}f_{0}=2f_{0}\ ,\nonumber
\end{equation}
such that
\begin{equation}
r=\ ^{12}\sqrt{2}\simeq1,0594631\ ,
\end{equation}
there is no exact frequency relation (\ref{con2}), but it is closely
approximated, in a compromise to favor the practice. The scale of equal
temperament is widely used as an universal tuning up of mostly popular musical
instruments, with a few exceptions as the violin or the singing natural human voice.

\section{Consonance Law of Sound Waves}

For the purpose to establish the law of sound consonance, the essential thing
is to know how two sound waves with different frequencies, $f_{1}$ and $f_{2}%
$, combine when produced simultaneously (harmony) or in a quick time sequence
(melody). Our daily hearing experience suggests us that the quality of the
sounds composition is essentially due to the frequencies combination of the
sounds, irrespective to their phases and amplitudes. In this sense, it is
sufficient, at least at the first sight, to examine just the time oscillation,
given by the trigonometric relation \cite{Tijonov}%

\begin{equation}
\cos2\pi f_{1}t+\cos2\pi f_{2}t=2\cos2\pi\frac{\left\vert f_{1}-f_{2}%
\right\vert }{2}t\cos2\pi\frac{(f_{1}+f_{2})}{2}t\ , \label{sum}%
\end{equation}
which shows a main wave with mean frequency
\begin{equation}
\overline{f}=f_{+}=\frac{(f_{1}+f_{2})}{2} \label{mean}%
\end{equation}
modulated by the beat frequency%

\begin{equation}
f_{-}=\frac{(f_{2}-f_{1})}{2}\text{ .} \label{beat}%
\end{equation}
Without loss of generality, we are going to suppose $f_{2}>f_{1}$.

We can see that, if
\begin{equation}
(f_{2}+f_{1})=n(f_{2}-f_{1})\ , \label{con}%
\end{equation}
that is,
\begin{equation}
\frac{f_{2}}{f_{1}}=\frac{(n+1)}{(n-1)}\ , \label{con1}%
\end{equation}
for integers $n>1$, the resulting wave is, yet, a regular and periodic wave,
behaving like a sound wave composed with harmonic sound components, as it is
in fact.

It can be seen inverting the trigonometric relation (\ref{sum}) above using
the frequency relation (\ref{con1}), which leads to
\begin{equation}
2\cos2\pi\frac{(f_{2}-f_{1})}{2}t\cos2\pi\frac{(f_{1}+f_{2})}{2}t=\cos
2\pi(n-1)f_{0}t+\cos2\pi(n+1)f_{0}t\ , \label{inverse}%
\end{equation}
for some frequency
\begin{equation}
f_{0}=f_{2}/(n+1)=f_{1}/(n-1) \label{fun}%
\end{equation}
used as the fundamental one. It is all we need to have an harmonious or
consonant combinations of sound waves. If the frequency relation (\ref{con1})
is an integer, the two sounds are harmonically related and trivially consonant
sounds. Otherwise, it defines the frequency relations in the range of an
octave, the frequency interval that comprises a musical scale, that is,
$f_{1}<f_{2}<$ $2f_{1}$, such that the relation (\ref{con1}) must be a
rational number limited by%
\begin{equation}
1<\frac{(n+1)}{(n-1)}<2\ ,
\end{equation}
for an integer $n>3$, remembering that $n=2$ and $n=3$ implies the harmonic
relations $f_{2}=3f_{1}$ and $f_{2}=2f$, respectively.

In equations (\ref{inverse}) and (\ref{fun}), it is assumed that%

\begin{equation}
f_{2}=(n+1)f_{0}%
\end{equation}
and%
\begin{equation}
f_{1}=(n-1)f_{0}\ ,
\end{equation}
which can be used in equations of\ the beat and the mean frequencies,
(\ref{beat}) and (\ref{mean}), respectively, resulting
\begin{equation}
f_{-}=\frac{(f_{2}-f_{1})}{2}=\frac{(n+1)f_{0}-(n-1)f_{0}}{2}=f_{0}
\label{beat1}%
\end{equation}
and%
\begin{equation}
\overline{f}=f_{+}=\frac{(f_{1}+f_{2})}{2}=nf_{0}\ , \label{mean1}%
\end{equation}
showing that, taking into account the physical consonance condition
(\ref{con}), all the relevant frequencies are related harmonically to a
fundamental one, $f_{0}$, which can assume values in the range%

\begin{equation}
0<f_{0}<f_{1}.
\end{equation}

It is easy to handle the situation when the primary sound waves has different
amplitude. Equation (\ref{sum}) must be replaced by\qquad%
\[
A_{1}\cos2\pi f_{1}t+A_{2}\cos2\pi f_{2}t=A_{1}\left[  \cos2\pi f_{1}%
t+\cos2\pi f_{2}t\right]  +\left(  A_{2}-A_{1}\right)  \cos2\pi f_{2}t
\]
and the condition of consonance applied to the equal amplitude terms, taken
aside the last term. Then, using the inverse relation (\ref{inverse}), the
last term can be reincorporated, resulting the more general formula%

\begin{equation}
A_{1}\cos2\pi f_{1}t+A_{2}\cos2\pi f_{2}t=A_{1}\cos2\pi(n-1)f_{0}t+A_{2}%
\cos2\pi(n+1)f_{0}t\ , \label{gen}%
\end{equation}
getting more confidence considering that in a real world a fine control of the
sound intensity is not a simple task.

An ultimate generalization is need to take into account the phase difference
between the primary sound waves. Such phase difference can be originated due
to the small, uncontrollable, time difference the primary sounds are produced.
Then, the left side of the sum (\ref{gen}) is better to be rewritten as%

\begin{equation}
A_{1}\cos2\pi f_{1}\left(  t-t_{1}\right)  +A_{2}\cos2\pi f_{2}\left(
t-t_{2}\right)  =W\left[  \cos\right]  +W\left[  \sin\right]  \ ,
\label{phase}%
\end{equation}
where%
\begin{equation}
W\left[  \cos\right]  =B_{1}\cos2\pi f_{1}t+B_{2}\cos2\pi f_{2}t \label{cos}%
\end{equation}
and%
\begin{equation}
W\left[  \sin\right]  =C_{1}\sin2\pi f_{1}t+C_{2}\sin2\pi f_{2}t \label{sin}%
\end{equation}
are the cosine and \ the sine wave components, respectively, with coefficients%
\begin{equation}
B_{1}=A_{1}\cos2\pi f_{1}t_{1},\ B_{2}=A_{2}\cos2\pi f_{2}t_{2}%
\end{equation}
of the cosine component and%
\begin{equation}
C_{1}=A_{1}\sin2\pi f_{1}t_{1},\ C_{2}=A_{2}\sin2\pi f_{2}t_{2}\
\end{equation}
of the sine component. Applying the consonance condition (\ref{con}), the
cosine component is just given by the equation (\ref{gen}),%

\begin{equation}
W\left[  \cos\right]  =B_{1}\cos2\pi(n-1)f_{0}t+B_{2}\cos2\pi(n+1)f_{0}t\ .
\label{cos1}%
\end{equation}

Using the trigonometric relation%
\[
\sin2\pi f_{1}t+\sin2\pi f_{2}t=2\cos2\pi\frac{\left(  f_{2}-f_{1}\right)
}{2}t\sin2\pi\frac{\left(  f_{2}+f_{1}\right)  }{2}t\ ,
\]
the same consonance condition (\ref{con}) works for the sine waves, resulting
\begin{equation}
W\left[  \sin\right]  =C_{1}\sin2\pi(n-1)f_{0}t+C_{2}\sin2\pi(n+1)f_{0}t\ .
\label{sin1}%
\end{equation}

It is possible to reverse all these proceeding, getting the quite general
expression
\begin{align}
&  A_{1}\cos2\pi f_{1}\left(  t-t_{1}\right)  +A_{2}\cos2\pi f_{2}\left(
t-t_{2}\right) \nonumber\\
& \nonumber\\
&  =A_{1}\cos2\pi(n-1)f_{0}\left(  t-t_{1}\right)  +A_{2}\cos2\pi
(n+1)f_{0}\left(  t-t_{2}\right)  t\ , \label{final}%
\end{align}
a guarantee that the consonance condition (\ref{con}) works in a real situation.

It is easy to verify that the frequency relations as $3:2:1$ and $4:5:6$
satisfy, all of than, the condition (\ref{con}) or (\ref{con1}). For example,
in the Pythagorean frequency relations, we have%

\begin{equation}
\frac{3}{2}=\frac{6}{4}=\frac{5+1}{5-1}\ ,\ \frac{3}{1}=\frac{2+1}%
{2-1},\ \frac{2}{1}=\frac{4}{2}=\frac{3+1}{3-1} \label{Pita}%
\end{equation}
and, in the Ptolomyan frequency relations,
\begin{equation}
\frac{6}{5}=\frac{12}{10}=\frac{11+1}{11-1},\ \frac{6}{4}=\frac{5+1}%
{5-1},\ \frac{5}{4}=\frac{10}{8}=\frac{9+1}{9-1}\ . \label{Ptolomy}%
\end{equation}

In Pythagorean Scale, in the table 1 there are 4 consonant relations (9/8,
4/3, 3/2 and 2) and in the table 2 there are 5 consonant relations 9/8. In the
Just Intonation Scale, in the table 3 there are 6 consonant relations (9/8,
5/4, 4/3, 3/2, 5/3, 2) and in the table 4 all of the 7 frequency relations are
consonant. It is the reason why the Just Intonation Scale is better than the
Pythagorean Scale.

Sound sources are vibrating systems, and produce sounds that are combinations
of a fundamental and its harmonics, with a particular combination defining the
timbre of the sound. So, to the consonance condition being consistent,
frequency relations like (\ref{con}) must be valid simultaneously to all, the
fundamental and its harmonic frequencies. Fortunately, it is so, as we can see
easily. Really, taking the compositions of all oscillating modes of the two
sound sources with fundamental frequencies $f_{1}$and $f_{2}$,
\begin{equation}
u_{1}(t)=\sum_{k=1}^{\infty}A_{k}\cos2\pi kf_{1}t \label{har1}%
\end{equation}
and
\begin{equation}
u_{2}(t)=\sum_{k=1}^{\infty}B_{k}\cos2\pi kf_{2}t\text{\ ,} \label{har2}%
\end{equation}
respectively, the combination of these two composed sounds (considering at a
moment the equal amplitudes $B_{k}=A_{k}$) becomes, from (\ref{sum}),
\begin{align}
u(t)  &  =\sum_{k=1}^{\infty}A_{k}\left(  \cos2\pi kf_{1}t+\cos2\pi
kf_{2}t\right) \nonumber\\
& \nonumber\\
&  =2\sum_{k=1}^{\infty}A_{k}\cos2\pi k\frac{\left\vert f_{1}-f_{2}\right\vert
}{2}t\cos2\pi k\frac{(f_{1}+f_{2})}{2}t\ . \label{sum1}%
\end{align}

Applying the consonance condition (\ref{con1}), supposing $f_{1}<f_{2}$, we
obtain the inverse trigonometric expansion given by (\ref{inverse}) in a
general form (\ref{gen}),%
\begin{equation}
u(t)=\sum_{k=1}^{\infty}A_{k}\cos2\pi k(n-1)f_{0}t+B_{k}\cos2\pi k(n+1)f_{0}t,
\label{gen4}%
\end{equation}
which is an harmonic series. Again, it is all we need to have an harmonious or
consonant combinations of sound waves.

\section{\smallskip Hearing Dissonance}

The physical consonance condition given by equation (\ref{con}) assures the
harmonic structure of the sound waves combination such that it behaves like a
natural sound waves with an enriched timbre. However, we have to take into
account the hearing sensibility of the human auditory system \cite{Augustus}
and \cite{Guyton}, able to recognize sounds at frequencies range from about
$20Hz$ to $20,000Hz$. Figure 1 Figure 1 shows a mathematical representation of
an artificial computer generated pure sound frequencies (a) $440Hz$ of the
musical scale standard $A_{4}$ note, (b) $264Hz$ corresponding to the note
$C_{4}$, (c) $495Hz$ of the note $B_{4}$ and (d) $20Hz$, the low audio
frequency threshold, in the time interval $0<t<0.05s$.

\bigskip

\begin{center}
Figure 1
\end{center}

\bigskip

If the beat frequency is bellow the low audio frequency threshold,
\begin{equation}
f_{0}=f_{-}\lesssim20Hz\text{ ,} \label{low}%
\end{equation}
that occurs when the primary frequencies $f_{2}$ and $f_{1}$ are close, which
implies big $n$, the fundamental frequency $f_{0}$ is going to be missing for
our audition. In this situation, the sound composition given by (\ref{sum})
cannot be heart as an harmonically related sounds (\ref{inverse}), but
instead, it is listen as%

\begin{equation}
\cos2\pi f_{1}t+\cos2\pi f_{2}t=\left(  2\cos2\pi f_{0}t\right)  \cos2\pi
nf_{0}t\ , \label{sum2}%
\end{equation}
an unique sound with frequency $nf_{0}$ modulated by an inaudible beat
frequency $f_{-}=f_{0}$. This modulation, for a very close frequencies, leads
to the well known beat phenomenon. Beat frequency near the transition region
between audible and inaudible frequencies might be confusing to the auditory
system like an antenna trying to tune in a signal with frequency in the border
of its work range frequencies. In such a way, the missing of the fundamental
frequency, while of physiological nature, can breaks down the consonance
condition of the primary frequencies. So, even satisfying the physical
consonance condition given by equation (\ref{con}), our audition will not
going to perceive as consonant. In the central octave frequencies range,
taking the first note, $C_{4}$, with frequency $f_{1}=264Hz$, the low
frequency limit (\ref{low}) occurs at the condition $n\gtrsim12$, above which
the auditive perception of consonance is going to be broken. As a result, the
major second one tone interval as the $C_{4}-D_{4}$, related by frequencies
ratio $9/8$, which corresponds to $n=17$, is not considered as consonant. With
primary frequencies $264Hz$ and $297Hz$ of the notes $C_{4}$ and $D_{4}$, the
resultant mean and the beat frequencies are $f_{+}=280.5Hz$ and $f_{-}%
=16.5Hz$, respectively. This beat frequency is out of the audible frequencies
range and, even satisfying the physical consonance condition (\ref{con}), this
is not considered as consonant. It is shown in the figure 2a, with the
presence of a characteristic wave modulation. In sequence, figure 2b is an
illustration of the sound composition with frequencies ratio $16/15$, the half
tone interval, satisfying the consonance condition with $n=30$. Frequencies
considered are $264Hz$ and $281.6Hz$ of the notes $C_{4}$ of the sharp
$^{\#}C_{4}$, resulting a sound with the mean frequency $f_{+}=272.8Hz$
modulated by the beat frequency $f_{-}=6.6Hz$. Figure 2c is a composition of
the notes $C_{4}$ ($264.0Hz$) and $B_{4}$ ($495.0Hz$), with frequencies ratio
$15/8$, a clearly non consonant relation. Figure 2d is a simple example of a
non consonant sound combination, with arbitrary frequencies, actually $264Hz$
($C_{4}$) and $340Hz$, mean and beat frequencies $f_{+}=302Hz$ and
$f_{-}=38Hz$, respectively. Figures 2a and 2b, while satisfying the physical
consonance, has a well defined periodicity commanded by their beat
frequencies, but figures 2b and 2d, do not satisfying the consonance
condition, show a clearly non periodic time evolution, resulting an undefined
pitch of the resultant sounds.

\bigskip

\begin{center}
Figure 2
\end{center}

\bigskip

The sensibility of the human auditory system to the audio frequency range is
individual and depends on a several factors such as the age or the health, to
cite some of them. As a consequence, the human perception of the consonance as
the pleasant sensation it produces, while reinforced by a physical condition
as the equation (\ref{con}), involves as well the individual subjective feeling.

Also, the human hearing perception of consonance or dissonance is not
absolute, and a slight deviation around the consonance condition (\ref{con})
is not perceived by the human auditory system. It is the reason why the
musical scale of equal temperament is acceptable.

In sequence, the table 5 shows the notes and respective frequencies of the
central octave in the Just Intonation Scale using the standard frequency
$f(A_{4})=440hz$ and the table 6 shows the beat frequencies of the two
adjacent notes.

\bigskip

\begin{center}%
\begin{tabular}
[c]{|c|c|c|c|c|c|c|c|}\hline
$C_{4}$ & $D_{4}$ & $E_{4}$ & $F_{4}$ & $G_{4}$ & $A_{4}$ & $B_{4}$ & $C_{5}%
$\\\hline
$264.0$ & $297.0$ & $330.0$ & $352.0$ & $396.0$ & $440$ & $495.0$ &
$528.0$\\\hline
\end{tabular}

Table 5: notes and frequencies of the central octave for the standard
$f(A_{4})=440hz$, in the Just Intonation Scale.

\bigskip%

\begin{tabular}
[c]{|c|c|c|c|c|c|c|}\hline
$D_{4}-C_{4}$ & $E_{4}-D_{4}$ & $F_{4}-E_{4}$ & $G_{4}-F_{4}$ & $A_{4}-G_{4}$
& $B_{4}-A_{4}$ & $C_{5}-B_{4}$\\\hline
$16.5$ & $16.5$ & $11.0$ & $22.0$ & $22.0$ & $27.5$ & $16.5$\\\hline
\end{tabular}

Table 6: beat frequencies of the pairs of adjacent notes of the central
octave, in the Just Intonation Scale.

\bigskip
\end{center}

Figure 3 contains typical consonant sound combinations, represented by the
$C_{4}-$ $E_{4}$ (major third) and the $C_{4}-$ $G_{4}$ (perfect fifth)
intervals, both satisfying the physical consonance condition (\ref{con}). In
(a), a sound composition \ of the notes $C_{4}$ ($264Hz$) and $E_{4}$ ($330Hz
$) frequencies ratio $5/4$, satisfying the consonance condition with $n=9$.
The resultant mean and beat frequencies are $f_{+}=297Hz$ and $f_{-}=33Hz$,
respectively. In (b), sound composition of the notes $C_{4}$ ($264Hz$) and
$G_{4}$ ($396Hz$), frequencies relation $3/2$, $n=5$. The resultant mean and
beat frequencies are $f_{+}=330Hz$ and $f_{-}=66Hz$, respectively. In (c) and
(d), the same major third $C_{4}-E_{4}$ and the perfect fifth $C_{4}-G_{4}$,
but with different relative phases. A principal characteristic of a consonant
combination of sounds is the well defined periodicity of the resultant wave, a
clear consequence of the condition of consonance (\ref{con}). Another
important feature is that the consonance condition is not affected by changing
the relative phases or the relative amplitudes of the composing sound waves.
An important consequence is that the consonance condition works as well as for
harmony and melody.

\begin{center}
\bigskip

Figure 3

\bigskip
\end{center}

\section{Conclusions}

An exact mathematical frequencies relation is presented to define a physical
consonance law of sound waves. It assures\ an harmonic structure of the sound
waves combination such that it behaves like a natural sound waves with an
enriched timbre, the beat frequency working as the fundamental one. It is not
affected by changing the relative phases or the relative amplitudes of the
primary sound waves and, as a consequence, the consonance condition is valid
for harmonic and melodic sound composition. Nevertheless, in situation where
the beat frequency is out of the audible range of frequencies encompassed by
the human auditory system, the consonance perception is going to be broken,
even the physical condition is satisfied. As a consequence, the human
perception of the consonance as the pleasant sensation it produces, while
reinforced by an exact physical relation, may involve as well the individual
subjective feeling that depends on a several factors such as the age or the
health, for example.

A principal characteristic of the consonance condition is the well defined
periodicity of the resultant wave. The dissonance is characterized by the
absence of periodicity in such a way that there is no defined pitch. It
suggests that the consonance condition should be released to a more weak form
of the consonance condition,%

\begin{equation}
(f_{2}+f_{1})=\frac{n}{m}(f_{2}-f_{1})\ ,\label{rel}%
\end{equation}
which implies%
\begin{equation}
\frac{f_{2}}{f_{1}}=\frac{(n+m)}{(n-m)}\ ,\label{rel1}%
\end{equation}
for integers $n$ and $m<n$, instead the more restrictive condition given by
equations (\ref{con}) and (\ref{con1}), from now on should be referred as the
strong form ($m=1$) of the consonance condition. In this released form, any
rational number is going to satisfy it for some integer $m$, but now the beat
frequency does not work as the fundamental one; they are related by
$f_{0}=f_{beat}/m$ and\ the chance of the long time periodicity given by
$1/f_{0}$ to be out of the hearing perception increases together $m$.

Anyway, first of all, sound consonance or dissonance is an human conception,
related to pleasant or unpleasant hearing sensation, which depends on the
physiology and the consequent acuity of the human auditory system and might be
strongly influenced by the cultural environment.

\bigskip

I'm thankful to Dr. Oscar Chavoya-Aceves for his important comment.

%

\newpage

\begin{center}
{\Huge Figure captions}
\end{center}

\bigskip

Figure 1: Mathematical representation of pure sound waves with frequencies (a)
$440Hz$ of the musical scale standard $A_{4}$ note, (b) $264Hz$ corresponding
to the note $C_{4}$, \ \ \ \ (c) $495Hz$ of the note $B_{4}$ and (d) $20Hz$,
the low audio frequency threshold.

\bigskip

Figure 2: Examples of non consonant sound combination. In (a), characteristic
beat modulation given by the one tone, major second, interval $C_{4}-$ $D_{4}%
$. (b) $16/15$ half tone $C_{4}$ ($264Hz$) and sharp $^{\#}C_{4}$ ($281.6Hz$)
interval , satisfying the consonance condition with $n=30$. The result is a
sound with the mean frequency $f_{+}=272.8Hz$ modulated by the beat frequency
$f_{-}=6.6Hz$. (c) $C_{4}$ ($264.0Hz$) and $B_{4}$ ($495.0Hz$) composition,
with frequencies ratio $15/8$, a clearly non consonant relation. (d) Simple
example of a non consonant sound combination, with arbitrary frequencies,
actually $264Hz$ ($C_{4}$) and $340Hz$, mean and beat frequencies
$f_{+}=302Hz$ and $f_{-}=38Hz$, respectively.

\bigskip

Figure 3: Typical examples of consonant sound waves combination. In (a), major
third $C_{4}-E_{4}$, with frequencies ratio $5/4$, $n=9$. In (b), perfect
fifth $C_{4}-G_{4}$, frequencies ratio $3/2$, $n=5$. In (c) and (d), the same
major third $C_{4}-E_{4}$ and the perfect fifth $C_{4}-G_{4}$, but with
different relative phases.


\begin{thebibliography}{9}                                                                                                %


\bibitem {Thomas}Thomas D. Rossing, \textit{The Science of Sound }(second
edition), Assison-Wesley, Reading (1990).

\bibitem {Feynman}Richard P. Feynman, Robert B. Leighton and Mattew Sands,
\textit{The Feynman Lectures on Physics}, Addison-Wesley, Reading (1963). (2002-03).

\bibitem {Harry}Harry F. Oslon, Music, \textit{Physics and Enginering} (second
edition), Dover, New York (1967).

\bibitem {Lapp}David Lapp, The Physics of Music and Musical Instruments, http://www.tufts.edu/as/wright\_center/physics\_2003\_wkshp/book.htm

\bibitem {Tijonov}A. N. Tijonov and A. A. Samarsky, \textit{Equaciones de la
Fisica Matematica}, MIR, Moscow (1972).

\bibitem {Augustus}Augustus L. Stanford, \textit{Foundations of Biophysics},
p. 124-145, Academic Press (1975).

\bibitem {Guyton}Arthur C. Guyton, \textit{Tratado de Fisiologia M\'{e}dica},
Ed. Guanabara Koogan, R.J. (1969).
\end{thebibliography}
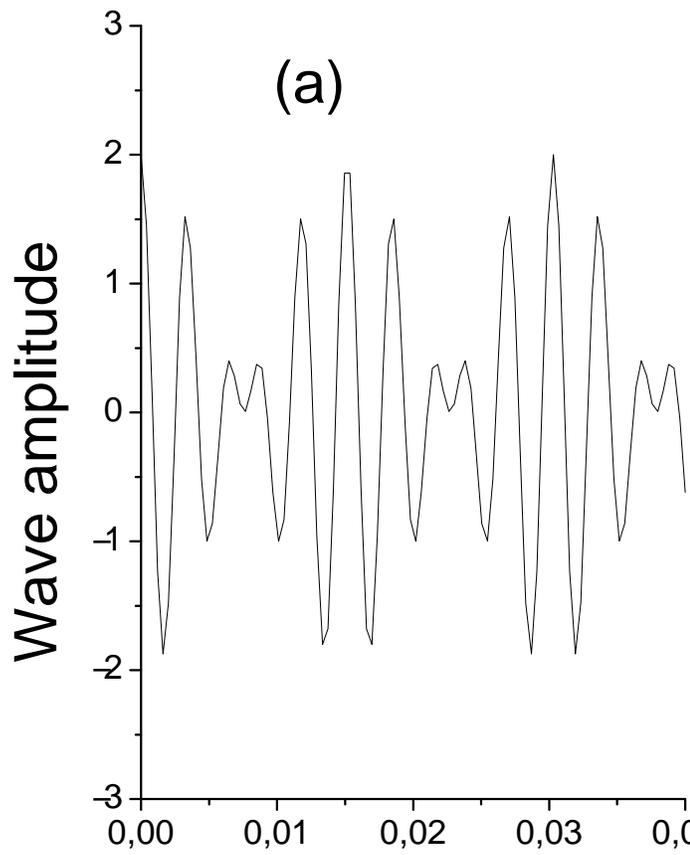
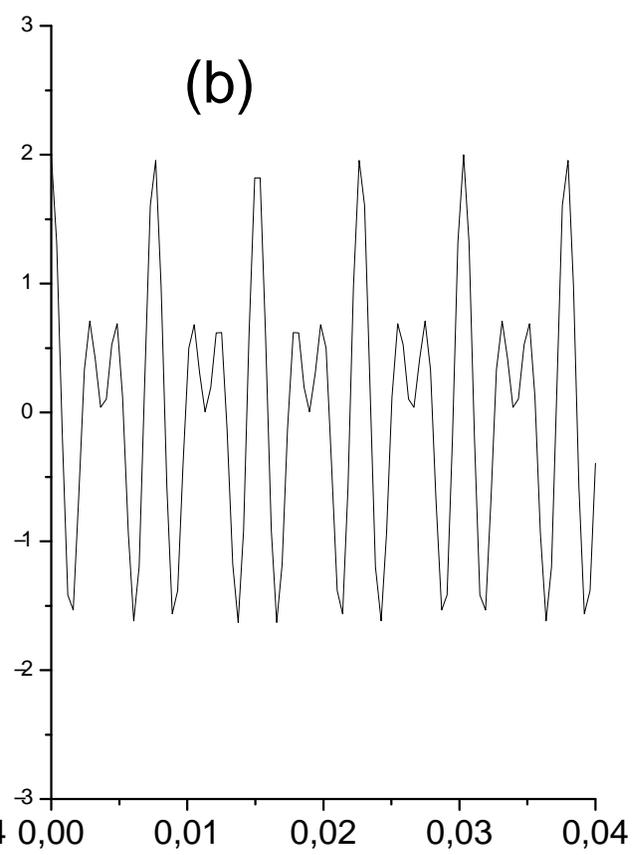
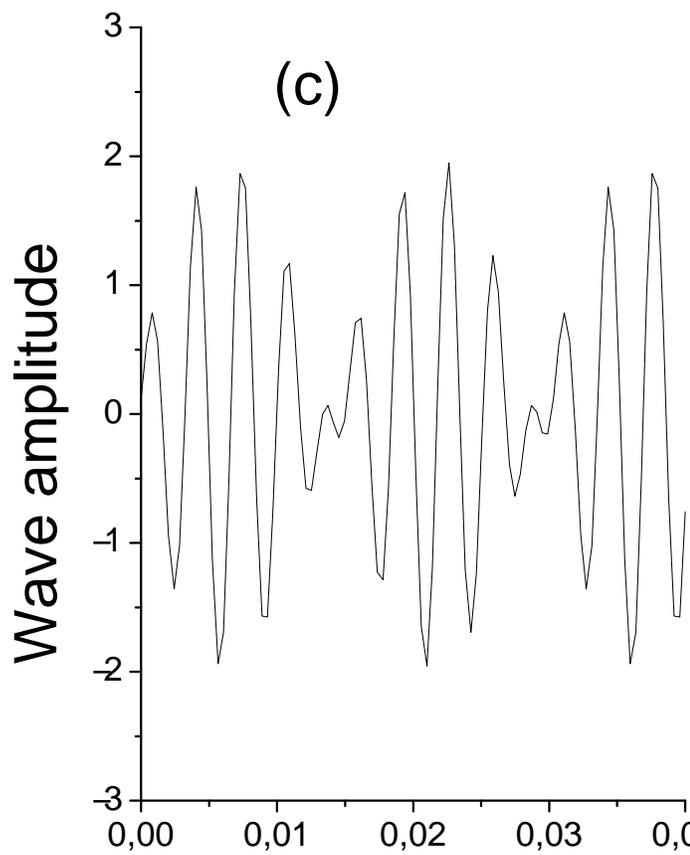
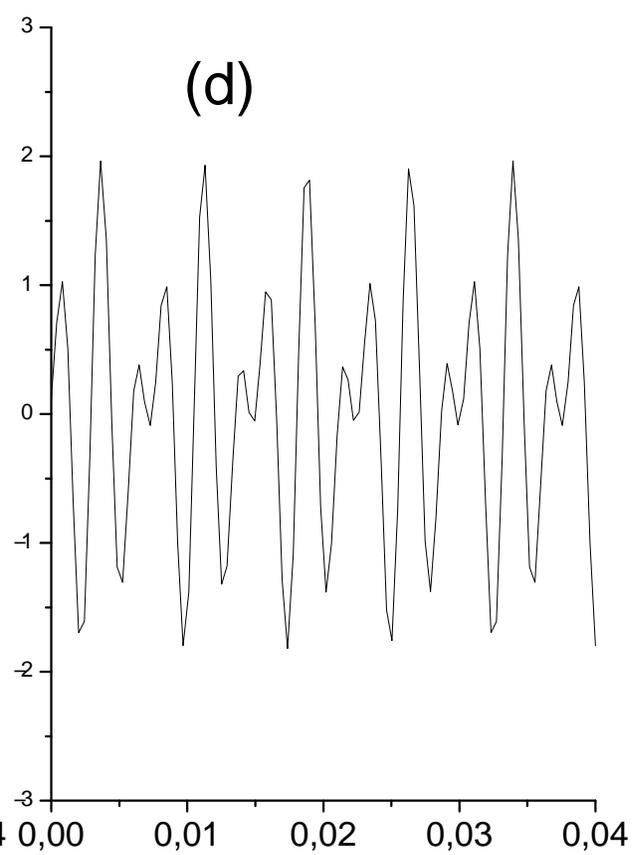
\end{document}